\documentclass[twocolumn,showpacs,prl]{revtex4}
\usepackage{graphicx}% Include figure files
\usepackage{dcolumn}% Align table columns on decimal point
\usepackage{bm}% bold math
\usepackage{amssymb}
\begin{document}

\title{A perspectival version of the modal interpretation of quantum mechanics and the origin of macroscopic behavior}

\author{Gyula Bene}
\email{bene@poe.elte.hu} \affiliation{Institute for Theoretical
Physics, E\"otv\"os University,
     P\'azm\'any P\'eter s\'et\'any 1/A, H-1117 Budapest, Hungary}
\author{Dennis Dieks}
\email{D.G.B.J.Dieks@phys.uu.nl} \affiliation{ Institute for the
History and Foundations of Science, Utrecht University, P.O.Box
80.000, 3508 TA Utrecht, \\The Netherlands }
\date{\today}

\begin{abstract}
We study the process of observation (measurement), within the
framework of a `perspectival' (`relational', `relative state')
version of the modal interpretation of quantum mechanics. We show
that if we assume certain features of discreteness and
determinism in the operation of the measuring device (which could
be a part of the observer's nerve system), this gives rise to
classical characteristics of the observed properties, in the
first place to spatial localization. We investigate to what
extent semi-classical behavior of the object system itself (as
opposed to the observational system) is needed for the emergence
of classicality. Decoherence is an essential element in the
mechanism of observation that we assume, but it turns out that in
our approach no environment-induced decoherence on the level of
the object system is required for the emergence of classical
properties.
\end{abstract}

 \maketitle

Modal interpretations (see, e.g.,
\cite{bene,dieks&vermaas,bacciagaluppi2,bub,vermaas&dieks}) aim at
assigning {\it properties} (or {\it states} that represent these
properties in a one-to-one way) to physical systems on the basis
of the standard quantum mechanical formalism, though stripped from
the postulates that attribute a special role to measurements. The
motivation for introducing states that correspond to physical
properties is the wish to give {\it descriptions} of systems, and
thus to transcend the traditional interpretational framework in
which systems are only discussed in terms of possible measurement
results. The removal of the measurement postulates has the same
background. We want to treat measurements as ordinary physical
interactions, and measurement outcomes as properties of measuring
devices or displays, and thus to remove any mysterious aspects of
the concept of quantum measurement. As a first step towards this
goal we assume all time evolution in Hilbert space to be unitary,
so that there is no collapse of the wave function.

The modal approach based on these starting-points has proved to be
appealing and successful when applied to situations in which the
Hilbert spaces are finite and the number of dimensions is not too
large \cite{bacciagaluppi and hemmo}. However, in the case of
continuous model systems, like freely moving particles or harmonic
oscillators (assumed to be interacting with the environment), the
existing prescriptions are not guaranteed to lead to the expected
classical properties. In particular, one such model study
\cite{bacciagaluppi} has failed to demonstrate more or less sharp
localization of a particle in circumstances in which one would
expect a classical description to be applicable. (As for other
kind of difficulties - not to be considered here - see
\cite{Vermaas_no_go}.) The essence of the failure to produce
localization is not the continuity of the model, since the
difficulty persists in models in which the Hilbert space has a
finite but large dimensionality. This can be clearly seen in
computer simulations, in which one always works with finite
Hilbert spaces.

At present it is not clear whether in more realistic models the
situation will improve and classical properties will result.
Nevertheless, it seems to be worth considering - still within the
modal scheme - another possibility, namely, that the observed
classical properties are inevitable and generic consequences of
the observation itself, but need not be present in the absence of
observations. As already emphasized, we consider observations and
measurements as ordinary physical interactions between object
system and measuring device, and treat them quantum mechanically.

It is indeed a-priori not implausible that applying the modal
scheme to the perceptual system itself will lead to results that
are in accordance with experience. The reason is that our nerve
system has an inherent discreteness, both in its spatial
structure (cells) and in its functionality (a nerve cell either
fires or does not fire). It is this kind of discreteness which
seems to be needed to recover the expected classical alternatives
within a quantum mechanical treatment \cite{bacciagaluppi and
hemmo}. Here, we will make a detailed investigation of the
implications of such a discreteness in a simple model (which can
be conceived either as a model of a digital measuring device or
as a very crude model of a part of the nerve system). The result
is that according to this model an observer looking at an object
will see the object localized: {\it from the point of view of the
observer the object is localized}. However, it turns out that the
object is delocalized from a different perspective.

The wider question addressed in this paper is whether these (and
similar) results can be fitted into a consistent and satisfactory
picture. An object cannot be both localized and delocalized, so
that it seems that inconsistency threatens. However, we will
propose, and to some extent develop, an interpretational scheme
(a generalisation of the Dieks-Vermaas modal interpretation)
according to which it is {\em not} contradictory to assign such
seemingly conflicting properties to an object. In this
`perspectival' version of the modal interpretation properties of
physical systems have a {\em relational} character and are
defined with respect to another physical system that serves as a
reference system \cite{bene}. It is important to emphasize
already now that the core idea of this new conceptual scheme is
that the different descriptions, given from different
perspectives, are equally objective and all correspond to
physical reality. Because of the relational character of the
descriptions this involves no contradiction. A contradiction
would only arise if different descriptions would be given from
{\it one} perspective, or from compatible perspectives that can
be combined into one. This will not happen in the interpretation
that we will propose. Furthermore, we will show that different
observers observing the same  object will agree about the
results, just as in classical physics (provided, of course, that
the observations do not change the object).

We shall also consider the time evolution of a macroscopic object
and study when and why the classical description becomes
applicable. Finally, we discuss the role of environment induced
decoherence.

\section{A perspectival version of the modal interpretation}

The approach we are going to explain is closely related to the
Dieks-Vermaas version of the modal interpretation: the same type
of rules are used to assign properties to physical systems. But
instead of the usual treatment in which properties are supposed to
correspond to monadic predicates, we will propose an analysis
according to which properties have a relational character.

Physical systems of a given kind are described within a
characteristic Hilbert space; we will allow arbitrary Hilbert
spaces. In our perspectival approach the state of a physical
system $S$ (corresponding to physical characteristics of $S$)
needs the specification of a `reference system' $R$ with respect
to which the state is defined. This reference system is a larger
system, of which $S$ is a part. As already mentioned, we will
allow that one and the same system, at one and the same instant of
time, can have different states with respect to different
reference systems. However, the system will have one single state
with respect to any given reference system. This state of $S$ with
respect to $R$ will be denoted by $\hat \rho^S_R$. It is a density
matrix, i.e., a Hermitian operator acting on the Hilbert space of
$S$ that is positive semidefinite and has unit trace. In the
special case in which $R$ coincides with $S$ the state is in
general (i.e., if there is no degeneracy, see below) a
one-dimensional projector
\begin{eqnarray}
\hat \rho^S_S=|\psi_S><\psi_S|\label{g3}
\end{eqnarray}
This state $\hat \rho^S_S$ (or equivalently $|\psi_S>$), the
`state of $S$ with respect to itself', is the same as `the
physical state' assigned to $S$ in the Dieks-Vermaas version of
the modal interpretation; i.e.\ it is one of the projectors
occurring in the spectral decomposition of the reduced density
operator of $S$, and $|\psi_S>$ is one of the eigenvectors, if
there is no degeneracy---see \cite{dieksy} for these ideas.

The rules for determining all states, for arbitrary $S$ and $R$,
are as follows. If $U$ is the whole universe, then $\hat \rho^U_U$
is taken as the quantum state assigned to $U$ by standard quantum
theory. If system $S$ is contained in system $A$, the state $\hat
\rho^S_A$ is defined as the density operator that can be derived
from $\hat \rho^A_A$ by taking the partial trace over the degrees
of freedom in $A$ that do not pertain to $S$:
\begin{eqnarray}
\hat \rho^S_A={\rm Tr}_{A\setminus S}\;\hat \rho^A_A\label{g4}
\end{eqnarray}
Any relational state of a system with respect to a bigger system
containing it can be derived by means of Eq.(\ref{g4}).

We already saw that for an arbitrary system $S$, contained in the
universe $U$, $\hat \rho^S_S$ is postulated to be one of the
projectors contained in the spectral resolution of $\hat
\rho^S_U$. If there is no degeneracy among the eigenvalues of
$\hat \rho^S_U$ these projectors are one-dimensional and the state
can be represented by a vector $|\psi_S>$, see Eq.(\ref{g3}); in
the case of degeneracy the state of the system with respect to
itself is a multi-dimensional projector. For simplicity we will in
the following focus on the non-degenerate case and assume that the
state of $S$ with respect to itself is given by one of the
eigenvectors $|\varphi^S_j>$ of $\hat \rho^S_U$.

The state $\hat \rho^U_U$ evolves unitarily in time. Because there
is no collapse of the wave function in our approach, this unitary
evolution of the total quantum state is the main dynamical
principle of the theory. Furthermore, we assume that the state
assigned to a \emph{closed} system $S$ undergoes a unitary time
evolution
\begin{eqnarray}
i\hbar\frac{\partial}{\partial t}\hat \rho^S_S=\left[\hat
H_S,\;\hat \rho^S_S\right]\label{g5}
\end{eqnarray}

As always in the modal interpretation, the theory specifies only
the probabilities of the various possibilities (the interpretation
is indeterministic): the probability that $|\psi_S>$ is the
eigenvector $|\varphi^S_j>$ is given by the corresponding
eigenvalue of $\hat \rho^S_U$. If the systems $S_1$, $S_2$, ...
$S_n$ are pair-wise disjoint and $U$ is the whole Universe, then
the joint probability that $|\psi_{S_1}>$ coincides with
$|\varphi^{S_1}_{j_1}>$, $|\psi_{S_2}>$ coincides with
$|\varphi^{S_2}_{j_2}>$,..., $|\psi_{S_n}>$ coincides with
$|\varphi^{S_n}_{j_n}>$, is given by
\begin{eqnarray}
P(j_1,j_2,...j_n)={\rm Tr} \left(\hat \rho^U_U \Pi_{i=1}^n
|\varphi^{S_i}_{j_i}><\varphi^{S_i}_{j_i}|\right) \label{g6}
\end{eqnarray}
We do not define joint probabilities if the systems are not
pair-wise disjoint. This is in accordance with the no-go theorem
by Vermaas \cite{Vermaas_no_go}. More generally, joint
probabilities cannot always be defined within the present approach
because states that are defined with respect to different quantum
reference systems need not be commensurable. In fact, this plays a
significant role in demonstrating that in our approach the
violation of Bell inequalities can be attributed solely to the
failure of the traditional concept of reality, and does not
involve nonlocality (see below, and \cite{Bene2}).

If a system and its complement are concerned, a simple calculation
based on the Schmidt representation of the state $|\psi_U>$ and
Eq.(\ref{g6}) shows that the states of $|\psi_A>$ and
$|\psi_{U\setminus A}>$ are uniquely correlated. (This result
played an important role in earlier versions of the modal
interpretation.) Therefore, knowledge of $|\psi_{U\setminus A}>$
is equivalent to knowledge of $|\psi_A>$. This suggests that one
may consider the state of $S$ with respect to the reference
system $A$, $\hat \rho^S_A$, alternatively as being defined {\it
from the perspective} $U\setminus A$ (here $A$ is an arbitrary
quantum reference system, while $U$ is again the whole universe).
Sometimes this concept of a `perspective' is intuitively more
appealing than the concept of a quantum reference system (cf.
\cite{Rovelli}). Nevertheless, there are some limitations
inherent in this alternative. First, if $A$ itself is the whole
universe, the concept of an external perspective cannot be
applied. Moreover, the state of the system $U\setminus A$ in
itself does not contain sufficient information to determine the
state of system $A$; one also needs the additional information
provided by $|\psi_U>$ in order to compute $|\psi_A>$. But
$|\psi_A>$ does contain all the information needed to calculate
$\hat \rho^S_A$ (cf. Eq.(\ref{g4})). We will therefore relativize
the states of $S$ to reference systems that contain $S$, although
we shall sometimes---in cases in which this is equivalent--- also
speak about the state of $S$ {\em from the perspective} of the
complement of the reference system.

Of course, we must address the question of the physical meaning of
the states $\hat \rho^S_A$. In our approach it is a fundamental
assumption that basic descriptions of the physical world have a
relational character, and therefore we cannot explain the
relational states by appealing to a definition in terms of more
basic, and more familiar, non-relational states. But we should at
the very least explain how these relational states connect to
actual experience. Minimally, the theory has to give an account of
what observers observe. We postulate that experience in this sense
is represented by the state of a part of the observer's perceptual
apparatus (the part characterized by a relevant indicator
variable, like the display in our simple model) {\em with respect
to itself}. More generally, the states of systems with respect to
themselves correspond to the (monadic)properties assigned by the
earlier, non-perspectival, version of the modal interpretation.

As we shall illustrate, the empirical meaning of many other
states can be understood and explained - by using the rules of
the interpretation - through their relation to these states of
observers, measuring devices, and other systems, with respect to
themselves.

\section{A model of the measurement}

The simple model we are going to use is sketched in Fig.1.
%\begin{figure*}[p]
\begin{figure}
\begin{center}
\includegraphics{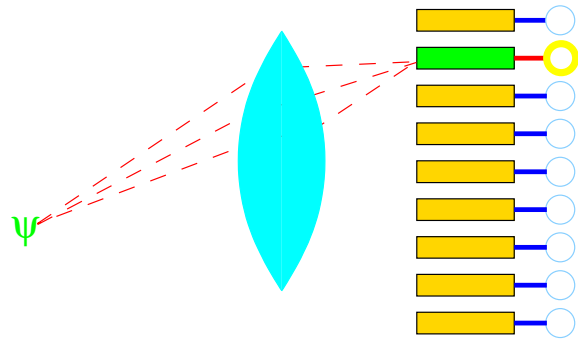}
\end{center}
\caption{Sketch of the model system}
\end{figure}
%\end{figure*}

The right hand side part of the drawing represents a digital
measuring device. It consists of several individual blocks that do
not interact with each other (as a representation of the situation
in the nerve system this is a strong simplification, though part
of the perceptual system can be modelled this way, especially in
situations in which the incoming stimulus has a low intensity).
The blocks are assumed to be very close to each other and their
diameters are larger than but comparable to the wavelength of
light. (The diameters of the rods and cones in the human retina
are 1-2 $\mu$m). Each block consists of a receptor (drawn as a
small rectangle) and a display (drawn as a circle). The operation
of a receptor may be described roughly by means of two states, one
corresponding to the receptor being excited, the other the
`ready-to-measure' state. We shall denote these by $|1>^R$ and
$|0>^R$, respectively. Any superposition of these states is also
allowed. A basis in the Hilbert space of the total measuring
device consisting of $N$ receptors is given by the states
${\Pi_{j=1}^N}|i_j>^R$, in which $'i_j'$, referring to the $j$-th
receptor, can be 0 or 1. The displays have corresponding states
that we denote by $|1>^D$ and $|0>^D$ for an individual display,
and by ${\Pi_{j=1}^N}|i_j>^D$ for the whole set of displays.

To make the ideas clear we shall assume that the interaction
between the receptors and displays is such that if the $j$-th
receptor is excited, and its state accordingly becomes $|1>^R$,
then the $j$-th display will with certainty end up in the
corresponding indicator state $|1>^D$. In other words, the
receptor-display system accomplishes an ideal von Neumann
measurement. This is known to be approximately realizable. In the
last section of this paper we will investigate a more refined form
of the model, in which the degeneracy of the just-mentioned
receptor and display states is taken into account.

The just-mentioned assumption immediately implies that an
arbitrary measurement leads to the entangled state
\begin{eqnarray}
\sum_{\{i_j\}} c_{i_1,i_2,...,i_N}|\Psi_{i_1,i_2,...,i_N}>\otimes
{\Pi_{j=1}^N}|i_j>^R\otimes|i_j>^D \label{e1}
\end{eqnarray}
of the whole model system, where the states
$|\Psi_{i_1,i_2,...,i_N}>$ of the measured object+environment
system are normed but not necessarily mutually orthogonal. Due to
the one-to-one relationship between the orthonormal states
${\Pi_{j=1}^N}|i_j>^R$ and ${\Pi_{j=1}^N}|i_j>^D$, the reduced
density matrix of the display system is
\begin{eqnarray}
\sum_{\{i_j\}} |c_{i_1,i_2,...,i_N}|^2 {\Pi_{j=1}^N}|i_j>^D\mbox{
}^D \mbox{\hspace{-0.2cm}}<i_j| \;,\label{e2}
\end{eqnarray}
i.e., it is diagonal in the `definite display result' basis. Its
eigenstates (except in case of degeneracy, i.e.\ exact equality of
the squared coefficients in (\ref{e2})) are the elements
${\Pi_{j=1}^N}|i_j>^D$ of the display result basis.

A basic principle of our interpretation is that the eigenstates of
the reduced density matrix are the `states of the system with
respect to itself' and correspond to possible physical properties
of the system (in the same way as the states assigned in the
earlier versions of the modal interpretation did). One of these
possibilities will be actually realized, and the probability for
any particular eigenstate of representing the actual state of
affairs is given by the value of the corresponding eigenvalue of
the reduced density matrix. It can thus be concluded from
Eq.(\ref{e2}) that the observation has a definite result
corresponding to one of the display states. As stressed before,
in the perspectival approach this definite property is
represented by the state of the display with respect to
itself---this leaves it open that the state and the corresponding
properties may be different if relativized to another reference
system. Below we will indeed encounter an example of such a
difference in state ascriptions corresponding to different
perspectives.

At this point we can already make a brief remark about the role
of decoherence. In the model the environment of the display
system is the receptor system, while coupling to the rest of the
world is neglected. Therefore, environment-induced decoherence in
the usual sense does not play a role here, although entanglement
between receptors and displays is essential. In other words, the
display states are `decohered' by their correlation with the
mutually orthogonal receptor states.

That a measurement, performed with a given device, invariably
leads to one of a number of alternatives that are determined by
the nature of the device and are independent of the measured
object (the latter determines only the probability that a
particular outcome occurs) has always been a standard assumption
in quantum measurement theory. In the present treatment we derive
this assumption from the modal approach applied to our specific
model of the measuring device. It should be noted that we did not
presuppose classical features of the device; the whole model is
treated quantum mechanically.

Suppose that a single photon is scattered from the object. If the
object possesses a large mass, we may neglect the back reaction
(recoil), so the total wave function of object and photon after
the scattering is of the form
\begin{eqnarray}
\int dx \int dx_e \int dx_{ph}\Psi(x,x_e)\varphi(x_{ph},x)\nonumber\\
\times |x>\otimes|x_e>\otimes|x_{ph}> \label{e3}
\end{eqnarray}
In this equation $x$ stands for the position of the object,
whereas $ x_{e}$ and $x_{ph}$ denote the degrees of freedom of
the environment and the photon, respectively. After the photon
has been absorbed in one of the receptors, the state of the
complete system (including the receptors and the displays) is
\begin{eqnarray}
\int dx \int dx_e \Psi(x,x_e)|x>\otimes|x_e>\nonumber\\
\otimes \sum_{j=1}^N c_j(x)\Pi_{i=1}^N|\delta_{i,j}>^R\otimes
|\delta_{i,j}>^D \label{e4}
\end{eqnarray}
In this equation $c_j(x)$ is the amplitude that the photon which
has been scattered from $x$, in the situation depicted in Fig.\ 1,
is absorbed at a later time in the $j$-th receptor. The amplitude
is negligible unless the position of the $j$-th receptor is near
the geometrical optical image of the point $x$. It follows that
the reduced density matrix of the display system is given by
\begin{eqnarray}
\sum_{j=1}^N \int dx
\rho(x,x)|c_j(x)|^2\Pi_{i=1}^N|\delta_{i,j}>^D<\delta_{i,j}|^D\label{e5}
\end{eqnarray}
Here
\begin{eqnarray}
\rho(x,x')=\int dx_e \Psi(x,x_e)\Psi^*(x',x_e)\label{e6}
\end{eqnarray}
stands for the reduced density matrix of the object in coordinate
representation, still before the object is exposed to the light.
Due to the recoil-free nature of the interaction, the diagonal
elements $\rho(x,x)$ are not influenced by the light scattering
(compare Eq.(\ref{e4prime}) below). Therefore, according to the
modal scheme the actual physical condition of the display system
(its state with respect to itself) is
\begin{eqnarray}
{\Pi_{i=1}^N}|\delta_{i,j}>^D \label{e7}
\end{eqnarray}
with probability
\begin{eqnarray}
\int dx \rho(x,x)|c_j(x)|^2\label{e8}
\end{eqnarray}
Note that Eq.\ (\ref{e5}) is a special case of Eq.\ (\ref{e2}),
thus Eq.\ (\ref{e7}) actually follows from the previous general
considerations.

In a more general treatment, we should also take into account
that the particle experiences a back-reaction because of its
interaction with the photon. Instead of Eq.(\ref{e4}), we then
have
\begin{eqnarray}
\int dx \int dx_e \Psi(x,x_e)|\xi_x>\otimes|x_e>\nonumber\\
\otimes \sum_{j=1}^N c_j(x)\Pi_{i=1}^N|\delta_{i,j}>^R\otimes
|\delta_{i,j}>^D\label{e4prime}
\end{eqnarray}
Here the states into which the position states $|x>$ are
transformed by the recoil are denoted by $|\xi_x>$; these states
are not necessarily mutually orthogonal. The probability of the
j-th display being excited now becomes
\begin{eqnarray}
\left|\left|\int dx \int dx_e \Psi(x,x_e)\;c_j(x)|\xi_x>
\otimes|x_e>\right|\right|^2
\nonumber\\
=\int dx \int dx' \rho(x,x')\;c_j(x)c^*_j(x')<\xi_{x'}|\xi_x>
\end{eqnarray}
Since the states $|\xi_x>$ are in general not mutually
orthogonal, non-diagonal elements of the reduced density matrix of
the object enter the expression. If the recoil is negligible, one
has $|\xi_x>=|x>$ and Eq.(\ref{e8}) is recovered.

It is instructive to calculate the reduced density matrix $\tilde
\rho(x,x')$ of the object (state of the object with respect to the
whole system) after it has been exposed to light, but still
before the absorption of the photon in the receptors. Using Eq.\
(\ref{e3}) one obtains
\begin{eqnarray}
\tilde \rho(x,x')= \rho(x,x')\int dx_{ph} \varphi(x_{ph},x)
\varphi^*(x_{ph},x')\label{e9}
\end{eqnarray}
After the absorption of the photon in the receptors one gets the
same result because the time evolution during the absorption is
unitary and the object degrees of freedom are not involved in the
interaction. A calculation based on Eq.\ (\ref{e4}) gives the
alternative expression
\begin{eqnarray}
\tilde \rho(x,x')= \rho(x,x')\sum_{j=1}^N
c_j(x)c_j^*(x')\label{e10}
\end{eqnarray}
The equality of expressions (\ref{e9}) and (\ref{e10}) gives a
condition the `transfer functions' $c_j(x)$ must satisfy.

According to the modal interpretations one of the eigenstates of
$\tilde \rho(x,x')$ represents the state of the object with
respect to itself (this is the physical state of the object
according to the terminology of the Dieks-Vermaas modal
interpretation). In general, these eigenstates are not localized
\cite{bacciagaluppi}, so they do not correlate with the result of
the observation, which indicates a definite position.

The question naturally arises what state the observation then
corresponds to. According to the modal scheme the answer follows
from the biorthogonal decomposition of the state of the whole
system (note that Eq.(\ref{e4}) is already of this form). Thus,
the observation is perfectly correlated to the state of the
object+environment+receptor system with respect to itself. In
other words, from the perspective of $D$ all information about the
object state is contained in the relational state of the object
with respect to object+environment+receptor. Direct calculation
based on Eq.(\ref{e4}) yields for this object state from the
perspective of $D$:
\begin{eqnarray}
\rho^{O}_{U\setminus D}(x,x')= \frac{c_j(x)\rho(x,x')c_j^*(x')}{
\int dx |c_j(x)|^2\rho(x,x)}\label{e11}
\end{eqnarray}
In this equation the superscript $O$ refers to the object, and the
subscript $U\setminus D=O+E+R$ to the reference system, which is
the complement of the system $D$ defining the perspective.

Owing to the properties of the amplitudes $c_j(x)$, for a given
$j$ the expression (\ref{e11}) will be appreciable only if $x$ and
$x'$ are located in a small interval (from which light will be
scattered to the $j$-th receptor). This means that the object is
indeed localized from the point of view of $D$ (or, equivalently,
with respect to---considered as part of---$O+E+R$). It should be
noted that the form of the state of $O+E+R$ is determined by the
interaction constituting the observation that took place. Without
the observation we could not speak of localization as a relational
property of the object.

The example illustrates how different properties may be ascribable
to an object from different perspectives. In this case, the state
of the object with respect to itself is not localized. However, if
the complement of the display system is chosen as the reference
system, a description in terms of a localized state does apply.
Both descriptions are objective, but relational; they involve the
specification of different reference systems. An analogy may be
helpful to see that the relational character of descriptions does
not entail a lack of objectivity. According to the special theory
of relativity one and the same object can be described as moving
or as resting, without the implication that one of these
descriptions is more fundamental than the other. Both descriptions
are entitled to be called objective, but make use of different
`perspectives'. The ascription of properties in our interpretation
of quantum mechanics has a similar relational character.

We will now consider a situation in which two or more
measurements are performed on the same object, by means of two
different measuring devices. We know from everyday experience
that the results of such measurements will (more or less) agree.
Our traditional notions concerning physical reality (in
particular, the idea that the properties of objects are
independent of any perspective and can be represented by
non-relational, monadic, predicates) are to a large extent based
on this and similar facts of experience. It is therefore important
to show that our perspectival approach is in accordance with this
agreement among observers (or measuring devices).

For simplicity we assume that the two devices are at the same
position and do not disturb each other. In this way we ensure
that both measurements take place under identical circumstances.
Analogously to Eq.(\ref{e4}), the state of the compound system
consisting of object, first device, and second device will after
the measurement be given by
\begin{eqnarray}
\int dx \int dx_e \Psi(x,x_e)|x>\otimes|x_e>\nonumber\\
\otimes \sum_{j=1}^N c_j(x)\Pi_{i=1}^N|\delta_{i,j}>^{R_1}\otimes
|\delta_{i,j}>^{D_1}\nonumber\\
\otimes \sum_{k=1}^N c_k(x)\Pi_{l=1}^N|\delta_{l,k}>^{R_2}\otimes
|\delta_{l,k}>^{D_2} \label{f4}
\end{eqnarray}
If we determine from this the state of the first display system
with respect to itself, we find
\begin{eqnarray}
{\Pi_{i=1}^N}|\delta_{i,j}>^{D_1} \label{f7}
\end{eqnarray}
with probability
\begin{eqnarray}
\int dx \rho(x,x)|c_j(x)|^2\label{f8}
\end{eqnarray}
Similarly, the state of the second display system with respect to
itself is
\begin{eqnarray}
{\Pi_{l=1}^N}|\delta_{l,k}>^{D_2} \label{f7a}
\end{eqnarray}
with probability
\begin{eqnarray}
\int dx \rho(x,x)|c_k(x)|^2\label{f8a}
\end{eqnarray}
According to the standard rules of the modal interpretation
(\cite{bene,dieksx}, see Eq.(\ref{g6}) ) the joint probability
that the state of the first display system with respect to itself
is given by Eq.(\ref{f7}) and the state of the second display
system with respect to itself is given by Eq.(\ref{f7a}) is
\begin{eqnarray}
\int dx \rho(x,x)|c_j(x)|^2|c_k(x)|^2\label{f9}
\end{eqnarray}
This expression has the same form as the joint probability
distribution predicted by classical theory for a situation in
which independent measurements are made on each member of an
ensemble of systems distributed in space with probability density
$\rho(x,x)$. The properties of the coefficients $c_j(x)$ imply
that the expression (\ref{f9}) vanishes unless $j\approx k$.
Indeed, $c_j(x)$ is practically zero unless the $j$-th receptor
block is situated near (i.e., in a distance of few wavelengths) of
the geometrical optical image of $x$. Therefore, our
interpretational scheme predicts that two observers looking at the
same macroscopic object, at the same time and under identical
circumstances, will see it (practically) in the same place. We
assumed in the calculation that the interaction between the photon
and the object was recoil-free; this is justified in the case of a
macroscopic object. We have already seen that if recoil should be
taken into account, off-diagonal elements $\rho(x,x^{\prime})$
from the narrow band in the matrix where $c_j(x)$ and
$c_j(x^{\prime})$ are both appreciably different from zero will
enter the expression for the probability of the j-th receptor
being excited. Moreover, if there is substantial recoil the second
measurement will show a different result than the first, because
of the disturbance caused by the first measurement. This is not
different from what would happen according to classical physics.

In expression (\ref{f9}) it does not matter whether the system was
in a pure or mixed state prior to the measurement. Therefore, in
our analysis of the observational mechanism macroscopic objects
will be observed as localized quite independently of whether
decoherence of the object state by interaction with its
environment has taken place. Our analysis indicates that this
point generalizes to arbitrary observation mechanisms that possess
the characteristic finiteness and determinism assumed in our
model. This result brings to light an important difference between
our approach, in which features of the measurement process take a
central role, and approaches according to which the localization
of macroscopic objects is due to environment-induced decoherence
of the object state. As emphasized before, according to the modal
interpretation environment-induced decoherence will in general
{\it not} guarantee that a macroscopic object will be localized
(in the sense that its state with respect to itself is localized),
because of the lack of localization of the eigenstates of the
object's reduced density matrix.

A similar analysis applies to the case in which the measurement is
repeated, possibly many times and in rapid succession, by means of
the same device. An adequate mathematical treatment of that case
includes the description of a memory which stores the result of
the first measurement (this memory would be analogous to the first
display system in the above situation) and a mechanism which
resets the measuring device and prepares it for the next
measurement. Without discussing the details, we just mention that
the results (especially the counterpart of Eq.(\ref{f9})) are
completely analogous.

\section{The notion of physical reality in the perspectival
approach} Although the agreement between different observers,
which fits in naturally with the classical notion of physical
reality and may even seem to imply it, was just found to be
present in our interpretation of the quantum formalism as well,
the overall picture of physical reality that emerges is very
different from the usual one. A good starting-point for an
explanation of the differences is a discussion of the
applicability of the Einstein-Podolsky-Rosen reality criterion.
This criterion says that

{\em If, without in any way disturbing a system, we can predict
with certainty (i.e., with probability equal to unity) the value
of a physical quantity, then there exists an element of physical
reality corresponding to this physical quantity.}\cite{EPR}.

Let us see how this applies to the just-discussed measurement
situation, in which the object interacts with the impinging
photons, which is followed by an interaction between the scattered
photons and the receptors, after which there finally is an
interaction between the displays and the receptors. The whole
process is repeated in the second measurement. The result shown
by the first display system allows an (almost) certain prediction
of the result of the second measurement. The question is what we
can say about the state of the object, after the light has been
scattered for the first time, on the basis of the EPR-criterion.

The interactions between the photons and the receptors and
between the receptors and the displays obviously do not disturb
the object, which may find itself at a large distance. As soon as
the display shows a result (or if the result is read off, but we
prefer to avoid the introduction of a conscious observer), there
is a one-to-one relation with the result of the second
measurement. In other words, the result of the second measurement
is predicted, with certainty, by the result of the first
measurement. As we have seen, the prediction is that the system
will be found at a definite position. At first sight, the
EPR-criterion therefore seems to imply that the object system
already possessed a definite position from the moment it
interacted with the photons.

However, in the approach that we are explaining things are not so
straightforward. In our scheme, the physical quantity that is
predicted corresponds to a {\em relational} state of the object,
namely its state with respect to the object+environment+receptor
system (cf.\ Eq.(\ref{e11})). Now, the important point is that the
first measurement \emph{has given rise to the perspective} from
which it is possible to make this prediction. Therefore, although
it is true that there was no physical interaction between the
display and the object, the display nevertheless plays a part in
determining the object state with respect to
object+environment+receptor (which is the complement of the
display system itself). The physical interaction with the display
affected the reference system, and therefore influenced the
relational state.

The fact that the relevant states have this relational
(perspectival) character is responsible for the failure of
ordinary counterfactual reasoning: from the fact that no physical
disturbance has affected the object, it cannot always be concluded
that the object state has remained the same. One should also look
at the reference system, with respect to which the state is
defined, and see whether anything has changed \emph{there} that is
relevant.

Einstein, Podolsky and Rosen thought it very unlikely that the
quantities of an object system would depend on whether or not a
remote measurement is performed. Within the conceptual framework
of classical physics, in which properties attach to an object as
monadic (non-relational) predicates, this skepticism is completely
justified. However, in our present framework the possibility of
the dependence in question naturally appears, not as an effect of
physical disturbances acting on the object but as a consequence of
a change in the conditions that define the perspective. This
change comes about by local physical influences on the quantum
reference system.

This line of reasoning is in accordance with Bohr's qualitative
arguments that any reasonable definition of physical reality in
the realm of quantum phenomena should also include the
experimental setup \cite{Bohr}. In the present relational approach
states of a system are defined with respect to \emph{any} larger
physical system, so the concept of reality is not exclusively
connected to the presence of instruments. Nevertheless, in our
scheme too, observed reality contains elements relating to us as
observers in an essential way: we define the perspective. However,
this does not imply any subjectivism. The various relational
states follow unambiguously from the quantum formalism, and the
way the world should be described depends accordingly on objective
physical features (whether or not the observing system is
discrete, the nature of the interaction, whether the object has a
large mass, and so on).

These considerations show how the very concept of reality is
modified in our interpretation of quantum mechanics. The essential
new point is that quantum properties and quantum states possess a
relational character. In general, one may expect that this quantum
feature will not be noticeable on the level of observations,
because of the agreement between different observers. Yet, the
modification persists even in the macroscopic domain. As we saw,
the observed localization of macroscopic objects is absent from a
different perspective. And on the observational level, Bell-type
experiments reveal the untenability of the traditional notions of
reality (monadic properties combined with locality).

Let us return to the EPR reality criterion and draw conclusions
about its status within our conceptual framework. As it stands,
the criterion is ambiguous (as observed by Bohr in his reply to
Einstein, Podolsky and Rosen), since nothing is said about the
perspective from which the physical quantity whose value can be
predicted is defined. If the reference system {\it is} specified,
the criterion is valid if neither the described system nor the
quantum reference system is disturbed. That is, if it is possible
to predict a relational state without any changes either in the
reference system or the object, the state is there (as a part of
physical reality) independently of whether or not the prediction
is made.

Specifically, one could understand the EPR-criterion as referring
to the state of the system with respect to itself. In this case,
the well-known `no-signaling' theorem becomes relevant: a system's
density matrix (found by partial tracing from $\hat \rho^U_U$),
and therefore its eigenstates, will not change as a result of
things happening elsewhere (remember that we do not have collapses
of the wave function in our scheme). So, if it is possible to
predict the state of a far-away particle (w.r.t.\ itself) on the
basis of measurements performed elsewhere, we surely should
conclude that that state existed independently of those
measurements, and the EPR criterion therefore holds.

Let us now apply the EPR reality criterion to the case for which
it was devised, the case of distant correlated particles. We find
that the state of the second particle that becomes precisely known
after a measurement on the first particle, is the state of this
second particle with respect to the two-particle system (i.e., the
state from the perspective of the measuring device). However, it
cannot be concluded that this state was already present before the
measurement, because the state of the reference system w.r.t.\
itself (from which the state of particle 2 w.r.t.\ this reference
system is derived) was changed by the measurement. If one writes
down the states explicitly, applying the given rules to the
situations before and after the measurement, one easily
establishes that the state of particle 2 w.r.t.\ the two-particle
system indeed changes as a result of the measurement, in spite of
the fact that there was no mechanical disturbance of particle 2.

By contrast, the state of particle 2 with respect to itself does
not change and there is no influence on the reference system. So,
if the state of particle 2 with respect to itself can be predicted
from the result of the first measurement, application of the EPR
criterion is possible and yields a result which is in harmony with
quantum mechanics (in our interpretation): the state of particle 2
w.r.t.\ itself was indeed an element of physical reality already
before, and independently of, the measurement on particle 1. More
generally, although the EPR criterion can be upheld within our
conceptual framework (by specification of the missing reference
system), its application does not lead to the conclusion that
there are more elements of reality than the relational states
admitted in our interpretation from the outset.

As it appears, the modification of the reality concept proposed
here makes the introduction of `quantum nonlocality' superfluous.
Indeed, the change in the relational state of particle 2 (with
respect to the 2-particle system) can be understood as a
consequence of the local change in the reference system, brought
about by the measurement interaction. The local measurement
interaction is responsible for the creation of a new perspective
(the state of the measuring device), and from this new perspective
there is a new state of particle 2. This agrees with a conclusion
not infrequently drawn from the violation of Bell's inequalities,
namely that we should \emph{either} abandon the usual realism
concept (something we do here) \emph{or} give up the principle of
locality; but not necessarily both.

\section{Time evolution and correlations between measurements}
Let us now consider the case in which the two measurements take
place at different instants of time. As before, we shall assume
that the measurements are performed by two different measuring
devices, both situated at the same place. In this section we
suppose that the interaction between the object and its
environment is negligible and that the object initially has its
own wave function. Finally, we restrict our considerations to the
case in which the object moves in one spatial dimension. After the
first measurement the whole system evolves freely during a time
interval $t$. At the end of this interval the total state (i.e.,
the state of the compound system, object+first receptor+first
display, with respect to itself) can be written as
\begin{eqnarray}
\int dx \int dx' G_t(x,x')\Psi(x') \times |x>\nonumber\\ \otimes
\sum_{j=1}^N c_j(x')\Pi_{i=1}^N|\delta_{i,j}>^{R_1}\otimes
|\delta_{i,j}>^{D_1} \label{f10}
\end{eqnarray}
In this equation $G_t(x,x')$ is the propagator representing the
free evolution between the measurements.

After the second measurement has finished, the state of the total
system reads
\begin{eqnarray}
\int dx \int dx' G_t(x,x')\Psi(x') \times|x>\nonumber\\ \otimes
\sum_{j=1}^N c_j(x')\Pi_{i=1}^N|\delta_{i,j}>^{R_1}\otimes
|\delta_{i,j}>^{D_1} \nonumber\\\otimes \sum_{k=1}^N
c_k(x)\Pi_{l=1}^N|\delta_{l,k}>^{R_2}\otimes |\delta_{l,k}>^{D_2}
\label{f11}
\end{eqnarray}
According to the modal interpretation rules, the state of the
object with respect to the object plus receptor system (i.e., the
object state from the perspective of $D_1$ and $D_2$) is one of
the states
\begin{eqnarray}
|\Psi_{j,k}>=\frac{\int dx \int dx'
c_k(x)G_t(x,x')\Psi(x')c_j(x')|x> }{ \sqrt{\int dx\;|c_k(x)|^2
\left|\int dx' G_t(x,x')\Psi(x')c_j(x')\right|^2}} \label{f10a}
\end{eqnarray}
with the probability
\begin{eqnarray}
P_{j,k}=\int dx\;|c_k(x)|^2 \left|\int dx'
G_t(x,x')\Psi(x')c_j(x')\right|^2\label{f10c}
\end{eqnarray}
If the object system is macroscopic, by which we mean that the
action is large compared to $\hbar$, the state $|\Psi_{j,k}>$ is
well localized in both coordinate and momentum space. The former
follows directly from the narrowness of the function $c_k(x)$.
Note, however, that the width in coordinate space is comparable
to the wavelength of the light, so it is still very wide on the
scale of the de Broglie wavelength of the object. As for the
momentum, we can make use of the fact that for a macroscopic
object the propagator has the approximate form
\begin{eqnarray}
G_t(x,x')=A_t(x,x')\exp\left(i\frac{S_t(x,x')}{\hbar}\right)\label{f10d}
\end{eqnarray}
where $S_t(x,x')$ is the classical action as a function of the
time and the initial and final coordinate of the orbit, and the
function $A_t(x,x')$ is smooth at the scale of the de Broglie
wavelength. Inserting this into Eq.(\ref{f10a}) and calculating
the probability distribution of the momentum, one gets via saddle
point integration
\begin{eqnarray}
\left|<p|\Psi_{j,k}>\right|^2\mbox{\hspace{5cm}}\nonumber\\
\propto \left|\int dx'
c_k(x)\exp\left(i\frac{S_t(x,x')}{\hbar}\right)
\Psi(x')c_j(x')\right|^2\;.\label{f10e}
\end{eqnarray}
Here $x=x(p,x')$ stands for the solution of the equation
\begin{eqnarray}
p=\frac{\partial S_t(x,x')}{\partial x}\;. \label{f10f}
\end{eqnarray}
Eqs.(\ref{f10e}) and (\ref{f10f}) imply that, again due to the
narrowness of $c_j(x')$ and $c_k(x)$, the momentum is distributed
in a narrow range around the classical value $\partial
S_t/\partial x$. In other words, the two measurements define a
classical orbit.

The question may be asked whether further measurements will
confirm that the object will be near this orbit. In order to
investigate this, let us consider a third measurement that takes
place after the second one, after an elapsed time interval $t'$.
The wave function of the whole system, including the object and
the three measuring devices, is now
\begin{eqnarray}
\int dx \int dx'' \int dx' G_{t'}(x,x'') G_t(x'',x')\Psi(x')
\nonumber\\
\times|x> \otimes \sum_{j=1}^N
c_j(x')\Pi_{i=1}^N|\delta_{i,j}>^{R_1}\otimes |\delta_{i,j}>^{D_1}
\nonumber\\\otimes \sum_{k=1}^N
c_k(x'')\Pi_{l=1}^N|\delta_{l,k}>^{R_2}\otimes
|\delta_{l,k}>^{D_2} \nonumber\\\otimes \sum_{n=1}^N
c_n(x)\Pi_{m=1}^N|\delta_{m,n}>^{R_3}\otimes |\delta_{m,n}>^{D_3}.
\label{h1}
\end{eqnarray}
If we now calculate the conditional probability of getting the
$n$-th result at the third measurement (the $n$-th display
showing a result), given that the $j$-th and the $k$-th result had
been obtained at the first and the second measurement,
respectively, we get
\begin{eqnarray}
\int dx |c_n(x)|^2\left|\int
dx''G_{t'}(x,x'')<x''|\Psi_{j,k}>\right|^2 \label{f12}
\end{eqnarray}
As discussed above, $|\Psi_{j,k}>$ is a wave packet with fairly
well defined coordinate and momentum. Therefore, it evolves in
time in such a way that the expectation value of the coordinate
and the momentum obeys the classical equations of motion, as
stated by Ehrenfest's theorem. Hence, the conditional probability
(\ref{f12}) will be different from zero only if $c_n(x)$ is
nonzero near the classical trajectory at time $t'$. If $n$ does
not correspond (in the sense of optical imaging) to the end point
of this trajectory, the conditional probability (\ref{f12})
vanishes. This result demonstrates how the classical laws of
motion emerge from a purely quantum mechanical description. Note
that the interaction of the object with its environment certainly
influences the resulting classical equations (for example,
through the appearance of dissipative terms), but the emergence
of classical motion itself is independent of whether or not there
is environment-induced decoherence.

In summary, what this section shows is that object systems that
have a semi-classical propagator (action large compared to
$\hbar$) follow classical paths. The mechanism of observation, and
the fact that we consider object states defined from the
perspective of the displays, is essential here. Without these
ingredients we would have no guarantee that the object wave packet
is small, and no localization and definite path would therefore
result. Our considerations indeed demonstrate in the literal sense
Heisenberg's famous statement {\em Die ``Bahn'' ensteht erst
dadurch, da{\ss} wir sie beobachten}.\cite{Heisenberg}

\section{A `delocalized' measuring device}
A central idea of our approach is not to {\em assume} the
localization of physical objects, but to derive it as a result of
the measurement interaction. We should therefore take into account
the possibility that even the measuring device itself may be
delocalized, in the sense that its wavefunction is not narrow in
the position representation. In this section we consider what
happens in this case. For simplicity we assume again that both the
object and the measuring device move in one dimension, along
parallel lines. We consider two simultaneous measurements taking
place at the same spot. Under these circumstances the total wave
function is
\begin{eqnarray}
\int dx\; dy\; dx_e\; \Psi(x,y,x_e)|x>\otimes|y>\otimes|x_e>\nonumber\\
\otimes \sum_{j=1}^N
c_j(x-y)\Pi_{i=1}^N|\delta_{i,j}>^{R_1}\otimes
|\delta_{i,j}>^{D_1}\nonumber\\
\otimes \sum_{k=1}^N
c_k(x-y)\Pi_{l=1}^N|\delta_{l,k}>^{R_2}\otimes
|\delta_{l,k}>^{D_2}\;, \label{h1a}
\end{eqnarray}
In this formula $y$ represents the center of mass coordinate of
the measuring device. We have assumed in (\ref{h1a}) that the
interaction between the photons and the apparatus depends only on
their mutual distance (and not on absolute position); in other
words, that the interaction Hamiltonian is translationally
invariant. As before, we can conclude from the strict coupling
between receptors and displays that the states of the display
systems with respect to themselves are such that only one display
block is excited. The joint probability that the $j$-th block of
the first device and the $k$-th block of the second device are
excited is given by
\begin{eqnarray}
\int dx\; dy\; dx_e\; |\Psi(x,y,x_e)|^2|c_j(x-y)|^2|c_k(x-y)|^2
\;. \label{h2}
\end{eqnarray}
As the functions $c_j(x)$ are well localized in their arguments
around a coordinate value that depends on their indices,
Eq.(\ref{h2}) implies that $j\approx k$. Indeed, if $j$ and $k$
differ appreciably, for any value of $x-y$ at least one of the
functions $|c_j(x-y)|^2$ and $|c_k(x-y)|^2$ is zero, so that the
integral in Eq.(\ref{h2}) vanishes. Therefore, both measurements
find the object at the same place.

The next question is what the state of the outside world is that
corresponds to this well-defined position. In order to answer
this question we should calculate the state of the system
consisting of the object and the center of mass of the measuring
device, with respect to the bigger system that also contains the
receptors, because this state gives a description from the
perspective of the displays. Using the rules of our approach, we
get
\begin{eqnarray}
\int dx\; dy\; dx'\; dy'\; dx_e\; \Psi(x,y,x_e)\Psi^*(x',y',x_e)
\nonumber\\\times
c_j(x-y)c_k(x-y)c_j^*(x'-y')c_k^*(x'-y')\nonumber\\\times
|x><x'|\otimes|y><y'| \;. \label{h3}
\end{eqnarray}
Clearly, this state is narrow only in the coordinate difference
$x-y$. If we are interested in the state of the object alone, a
narrow state emerges if the object's position is defined relative
to the measuring device. In order to see this, one may use the
canonical transformation $(x,y)\rightarrow (x-y, (x+y)/2)$, and
calculate the trace of the state (\ref{h3}) over the coordinate
$X=(x+y)/2$. One finds
\begin{eqnarray}
\int d\tilde x\; d\tilde x'\; dX\; dx_e\; \Psi(X+\tilde
x/2,X-\tilde x/2,x_e)\nonumber\\\times \Psi^*(X+\tilde
x'/2,X-\tilde x'/2,x_e)\nonumber\\\times c_j(\tilde x)c_k(\tilde
x)c_j^*(\tilde x')c_k^*(\tilde x') |\tilde x><\tilde x'|
\;,\label{h4}
\end{eqnarray}
where $\tilde x=x-y$. So a localized object state is still
obtained, even if the measuring device by itself may be
delocalized. This state is relative in two different senses: (i)
as before, it describes the object in a relational way, namely
from the perspective of the display system, and (ii) it
characterizes the object by means of its position relative to the
measuring device (observer).

\section{Environment-induced decoherence and the definiteness and stability of experience}
As we have seen in the introduction, a great number of degrees of
freedom, together with a decohering environment, stand in the way
of localization. In such a situation the eigenstates of the
density matrix tend to be delocalized in coordinate and momentum
space \cite{bacciagaluppi}. This gave rise to the problem, within
the modal approach, of how the fact can be explained that
macroscopic objects seem to possess well-defined positions. Our
proposed answer depended on an analysis of the observation
process. This shift of attention from the object to the measuring
device made it irrelevant whether the eigenstates of the object's
density matrix are localized or not. Indeed, we found that in an
observation only the diagonal elements (in coordinate
representation) of the density matrix influence the result; the
other elements, though essential for a determination of the
eigenstates, play no role here.

Although the original localization problem is thus solved within
our approach, one may wonder whether a similar problem does not
re-emerge for the measuring device itself. As we have seen, no
requirement about \emph{localization} in coordinate and momentum
space need to be made for the receptor and display in our model.
However, it needs to be investigated how the states of the device
behave if there are very many microscopic degrees of freedom and
if there is interaction between the environment and the
receptor-display system. We would like to be justified in
expecting that no superpositions of states corresponding to
different measurement results will arise. Independently of
whether there is a one-to-one correspondence between the
receptors and the displays (as in our simple model of a
measurement interaction), we would like to have as a general
result that the displays show definite and stable results,
because in the final analysis such display readings correspond to
our experience.

We should therefore investigate the properties of a display
system in interaction with its environment. We shall illustrate
below how environment-induced decoherence (\cite{Kiefer_Joos},
\cite{Zurek1}) tends to lead to definiteness and stability of the
display readings, even if there are very many internal
microscopic degrees of freedom. It is important, though, that the
dimensionality of the environment's Hilbert space is also very
high.

Let us consider first a single display. Let us divide the Hilbert
space of this system into two orthogonal subspaces, corresponding
to the ready and the excited states, respectively. Let $|1,j>$ and
$|2,j>$ be orthonormed bases within these subspaces. We assume
that the interaction Hamiltonian does not involve transitions
between these two subspaces, so that it can be written as
\newpage
\begin{eqnarray}
\hat H_{int}=\sum_{j,k} |1,j><1,k|\hat B_{j,k}^{(1)}\nonumber\\
            +\sum_{j,k} |2,j><2,k|\hat B_{j,k}^{(2)}
\label{dx11}
\end{eqnarray}
The two terms of $\hat H_{int}$ commute and the two subspaces are
invariant under the action of $\hat H_{int}$ and the associated
evolution operator $\hat U_t$. The physical justification for
assuming this form of the interaction Hamiltonian is that this way
an initial `ready-to-measure' state of the display never evolves
into an excited state (or vica versa) due to an interaction with
the environment only (i.e., when no measurement is performed). An
initial state that is a product of a state of the display system
and an environment state, $|\phi> \otimes |\xi>$, evolves
as\pagebreak
\begin{widetext}
\begin{eqnarray}
|\Psi_{t}> =\sum_{l} c_l^{(1)}
\exp\left(-i\frac{t}{\hbar}\sum_{j,k} |1,j><1,k|\hat B_{j,k}^{(1)}\right)|1,l>\otimes |\xi>\nonumber\\
+c_l^{(2)} \exp\left(-i\frac{t}{\hbar}\sum_{j,k} |2,j><2,k|\hat
B_{j,k}^{(2)}\right)|2,l>\otimes |\xi> \label{dx12}
\end{eqnarray}
\end{widetext}
where $c_l^{(1)}=<1,l|\phi>$ and $c_l^{(2)}=<2,l|\phi>$. The
reduced density matrix of the display system becomes

\begin{eqnarray}
\hat \rho=\sum_{j}\sum_{k} |1,j>\rho_{j,k}^{11}<1,k|
+|2,j>\rho_{j,k}^{22}<2,k|\nonumber\\
+|1,j>\rho_{j,k}^{12}<2,k|+|2,j>\rho_{j,k}^{21}<1,k| \label{dx13}
\end{eqnarray}
We want to show that in case of a large environment this density
matrix soon becomes block diagonal, so that its eigenstates
belong to either the first or the second subspace. This means
that the state of the system with respect to itself (which is one
of the eigenstates of the density matrix) is either a ready
state, or an excited state, but never a superposition of both.

The elements of the off-diagonal blocks can be expressed as
\begin{eqnarray}
\rho_{j,k}^{12}=\rho_{k,j}^{21\dagger}
=\sum_{l,l'}c_l^{(1)}c_{l'}^{(2)*}<\xi|\hat G_{k,l'}^{(2)\dagger}
\hat G_{j,l}^{(1)}|\xi> \label{dx14}
\end{eqnarray}
where
\begin{eqnarray}
\hat G_{j,l}^{(n)} =\left[\exp\left(-i\frac{t}{\hbar}  \hat
\mathbb{B}^{(n)} \right)\right]_{j,l}\quad (n=1,2)\;. \label{dx15}
\end{eqnarray}
In Eq.(\ref{dx15}) $\hat \mathbb{B}^{(n)}$ stands for a matrix
whose $(j,k)$-th element is the operator $\hat B_{j,k}^{(n)}$.

Expression (\ref{dx14}) is the scalar product of the states
\begin{eqnarray}
\sum_l c_l^{(1)} \hat G_{j,l}^{(1)}|\xi> \label{dx16}
\end{eqnarray}
and
\begin{eqnarray}
\sum_l c_l^{(2)} \hat G_{k,l}^{(2)}|\xi> \label{dx17}
\end{eqnarray}
As the operators  $\hat B_{j,k}^{(1)}$ and $\hat B_{j,k}^{(2)}$
are in general quite different, states (\ref{dx16}) and
(\ref{dx17}) soon behave like two randomly chosen vectors in the
$N$-dimensional Hilbert space of the environment. The expectation
value of the modulus square of the scalar product of two such
random vectors is $2^{-(N-1)}$, thus
\begin{eqnarray}
|\rho_{j,k}^{12}|\approx 2^{-\frac{N-1}{2}}. \label{dx18}
\end{eqnarray}
Therefore, if $N$ is large, $\hat \rho$ becomes approximately
block diagonal in the basis $|n,j>$. Suppose that there is a very
large, but finite, number $K$ of such basis vectors. The
constraints that the eigenvalues must be positive and that they
add up to unity lead to a small level spacing if $K$ is large.
Assuming equidistant eigenvalues, the level spacing of $\hat
\rho$ is $2/K^2$ \footnote{Random matrices typically exhibit level
repulsion, i.e., a tendency to avoid degeneracies and to have
approximately equidistant levels.}. The ensuing closeness of the
eigenvalues implies that the eigenstates tend to be
superpositions of the basis states $|n,j>$. The elements of the
off-diagonal block must be much smaller than the level spacing in
order to avoid a mixing between the two subspaces (compare
\cite{bacciagaluppi and hemmo}). The requirement
$2^{-\frac{N-1}{2}}\ll 2/K^2$ is easily satisfied if the
dimensionality of the environment's Hilbert space is large. If it
is satisfied, the eigenstates of $\hat \rho$ have the stable
property of belonging to one or the other subspace (ready or
excited states, respectively). It is this property that
corresponds to a definite outcome of a measurement.

A generalization to the case of several ($n_R$) displays is
straightforward. In that case the system of interest contains all
the displays, and one has to assume $2^{n_R}$ subspaces within the
system's Hilbert space that are invariant under time evolution.
The above formalism applies with these amendments \footnote{A more
detailed quantitative discussion will be published elsewhere.}.
The conclusion is again that interaction with an environment whose
Hilbert space has a sufficiently high number of dimensions leads
to a block diagonal density matrix. This density matrix will have
eigenstates which for each display belong stably to one or the
other subspace. This corresponds to a well-defined excitation
pattern of the display system.

\begin{acknowledgments}
One of us (G.B.) acknowledges the support given by the NATO
Science Fellowship Program, grant \mbox{No. T 031 724}, and a
J\'anos Bolyai Research Fellowship. He also thanks for the
hospitality extended by the Institute for the History and
Foundations of Science, Faculty of Physics and Astronomy, Utrecht
University.
\end{acknowledgments}

\end{document}